# Microsecond Bias Polarity Switching Reveals Hidden Charge Dynamics at Halide Perovskite Interfaces

Marián Betušiak[1], Roman Grill[1*], Eduard Belas[1], Petr Praus[1], Mykola Brynza[1], Mariselvam Karuppaiya[1], Mahshid Ahmadi[2*], Jonghee Yang[3], Artem Musiienko[4*]

[1] *Institute of Physics, Faculty of Mathematics and Physics, Charles University, Ke Karlovu 5, CZ 121 16 Prague, Czech Republic*

[*]*roman.grill@matfyz.cuni.cz*

[2] *Joint Institute for Advanced Materials, Department of Materials Science and Engineering, University of Tennessee, Knoxville, TN 37996, USA.*

[*]*mahmadi3@utk.edu*

[3] *Department of Chemistry, Yonsei University, Seoul 03722, Republic of Korea*

[4] *Young Investigator Group, Robotized Optoelectronic Material and Photovoltaic Engineering, Helmholtz-Zentrum Berlin für Materialien und Energie (HZB), Berlin, Germany.*

[*]*artem.musiienko@helmholtz-berlin.de*

## Abstract

We present a time-domain technique based on rapid bias polarity switching (BiPS) to probe charge transport and near-surface defects in halide perovskite single crystals. The method exploits interfacial extraction barriers, which cause carrier accumulation and subsequent release after bias reversal. BiPS combines surface sensitivity (~200 nm–2 µm) with millimeter-scale reach, enabling reconstruction of internal field profiles, detection of bulk space charge down to ~$10^9$ cm$^{-3}$, and resolution of microsecond–millisecond trap dynamics. In our setup the surface-state detection limit is ~$10^9$ cm$^{-2}$, and could be further improved by optimized illumination and readout.

Applied to melt-grown CsPbBr$_3$ (Cr/Cr) and solution-grown MAPbBr$_3$ (Cr/SnO$_2$/Cr), BiPS reveals interfacial barriers that drive hole accumulation and defect filling. CsPbBr$_3$ shows long-lived space charge (~$3 \times 10^{11}$ cm$^{-3}$) and ~250× faster hole extraction than MAPbBr$_3$, while trap analysis yields capture times of 1–100 µs, detrapping times of 20 µs–3 ms, and activation energies of 300–500 meV.

BiPS thus provides direct access to buried interfacial processes, disentangles electronic and ionic contributions, and offers a robust platform for evaluating contacts and guiding defect engineering in perovskite optoelectronic devices.

I Introduction

Metal-halide perovskites have gained significant attention in the field of optoelectronics due to their exceptional properties, including high absorption coefficient, tunable bandgap, and long carrier diffusion lengths[1–5]. Despite these advantages, the stability of perovskite-based devices is often limited by defects at surface and interface[6–8]. These defects can act as traps and non-radiative recombination centers accumulating charge, thereby reducing device operational lifetime[9–12]. Studies have shown that the defect density at surfaces and interfaces can be several orders of magnitude higher than in the bulk material[12,13], emphasizing the critical role of surface in enhancing device efficiency and stability.

It was reported that hole trapping and accumulation leads to phase segregation in mixed-halide perovskites, a process that is further enhanced in the presence of hole extraction barriers[14,15]. These challenges underscore the critical need for precise experimental techniques capable of probing the structure, energetics, and transport properties of perovskite bulk, perovskite surface and interface with adjacent contact

Despite extensive research on perovskite-based optoelectronic devices such as photodetectors, photovoltaics, and sensors[16,17], the spatial distribution of electric field and defects across the full device thickness remains poorly understood due to limitations of current characterization methods.

Most experimental techniques probe either the surface or the bulk, but lack the depth resolution and field sensitivity required to capture internal charge dynamics. Scanning electron microscopy (SEM) and atomic force microscopy (AFM) provide information on surface morphology and grain boundaries[18–21], while X-ray photoelectron spectroscopy (XPS)[22–25] and energy-dispersive X-ray spectroscopy (EDX) yield stoichiometric data, yet none of these methods reveal how specific defects influence charge transport and recombination.

Photoluminescence (PL) spectroscopy enables the study of recombination pathways via PL decay[9,26–28] but it is limited to fast processes (pico- to microsecond timescales), and typically misses slower phenomena such as carrier trapping, detrapping, or long-lived charge accumulation (>10 µs), which are crucial for understanding device stability and electric field evolution.

More recently, drive-level capacitance profiling (DLCP) has been used to map the spatial and energetic distributions of trap states in completed perovskite devices, demonstrating, in principle, depth-resolved sensitivity to interfacial versus bulk defects. Ni et al.[10] showed spatially varying trap densities in single-crystal and polycrystalline cells and discussed practical limits set by nonplanar depletion edges and material heterogeneity, which blur the profiling distance. At the same time, subsequent analyses[13,29] emphasized that capacitance-based techniques detect only charges that measurably perturb the electrostatic potential; thus, for low net defect densities and thin absorbers, the response is dominated by geometric and injection capacitances and can yield U-shaped profiles even in defect-free devices. This sets a thickness-dependent detectability threshold ($\sim 1/L^2$ scaling) and explains why many reported DLCP/C–V densities cluster near the limit of what these methods can reliably resolve. In ionically conductive perovskites[30,31], DLCP—like other frequency-domain spectroscopies—further suffers from model-dependent ambiguities unless ion migration and substitution schemes are explicitly accounted for; otherwise, evaluated defect characteristics can be biased by the competing ionic transport channel[16]. Although Ni et al. mitigated hysteresis and used forward/backward scans to argue minimal ion-migration impact in their specific samples, careful control and validation remain essential for general use.

In this work, we present an experimental technique based on analysis of pulsed photocurrent transients with a short optical excitation (~100 µs) followed by a rapid Bias Polarity Switching (BiPS). Unlike DLCP, BiPS senses current induced by carriers released from surface defects rather than the electrostatic imprint of trapped charge, so its sensitivity to surface states is not limited by absorber thickness.

This approach allows the identification of interfacial barriers as well as the evaluation of surface defect parameters such as energy levels and concentrations (down to $10^9$ cm$^{-2}$) and extraction rate of charge carriers through the barrier. While the time- and bias-dependent photocurrent induced by accumulated free carriers allows us to reconstruct the internal electric-field profile, quantify bulk space-charge densities down to ~$10^9$ cm$^{-3}$ in 1.5-mm crystals and determine the drift mobility. Because BiPS operates on photocurrent, ionic drift contributions can be separated (and the field-profile evolution directly monitored), enabling a diagnostic of whether ion migration is occurring. By tracking how the field profile and trap occupancy evolve after photoexcitation under controlled bias, BiPS cleanly differentiates bulk versus interfacial transport phenomena—without invoking the depletion-approximation assumptions that underpin DLCP.

## II. Surface to bulk electro-transport properties detected rapid bias polarity switching

To overcome the limitations of existing techniques in detecting the charge dynamics, electric field variations and space charge build-up, we introduce a novel experimental approach that sensitively tracks charge accumulation at buried interface. This method exploits the accumulation of photogenerated holes beneath a contact barrier — such as a Schottky barrier or insulating interlayer as an internal probe of interfacial transport dynamics. The principle of BiPS method is illustrated in Fig. 1a. When the sample is illuminated, photogenerated carriers are created, and holes drift under the influence of the electric field toward the cathode. A small fraction of these holes is able to cross the contact barrier via (i) thermal excitation and tunneling, (ii) hopping through localized states within the barrier, or (iii) recombining with injected electrons. However, most of the holes become trapped at the interface, either as free carriers or captured in shallow and deep surface trap states, leading to measurable charge accumulation. Prior to bias switching, the photocurrent follows the optical pulse, where the width of the rising (Fig. 1b, interval –6 $t_{tr}$ to –5 $t_{tr}$) and falling (–2 $t_{tr}$ to – $t_{tr}$) edges corresponds to the carrier transit time $t_{tr}$, i.e., the time required for photogenerated carriers to drift across the sample. In some cases, this signal can exhibit a tail resulting from de-/trapping of photocarriers in the bulk. However, the bulk de-/trapping is negligible in our samples and is therefore not displayed in Fig. 1b.

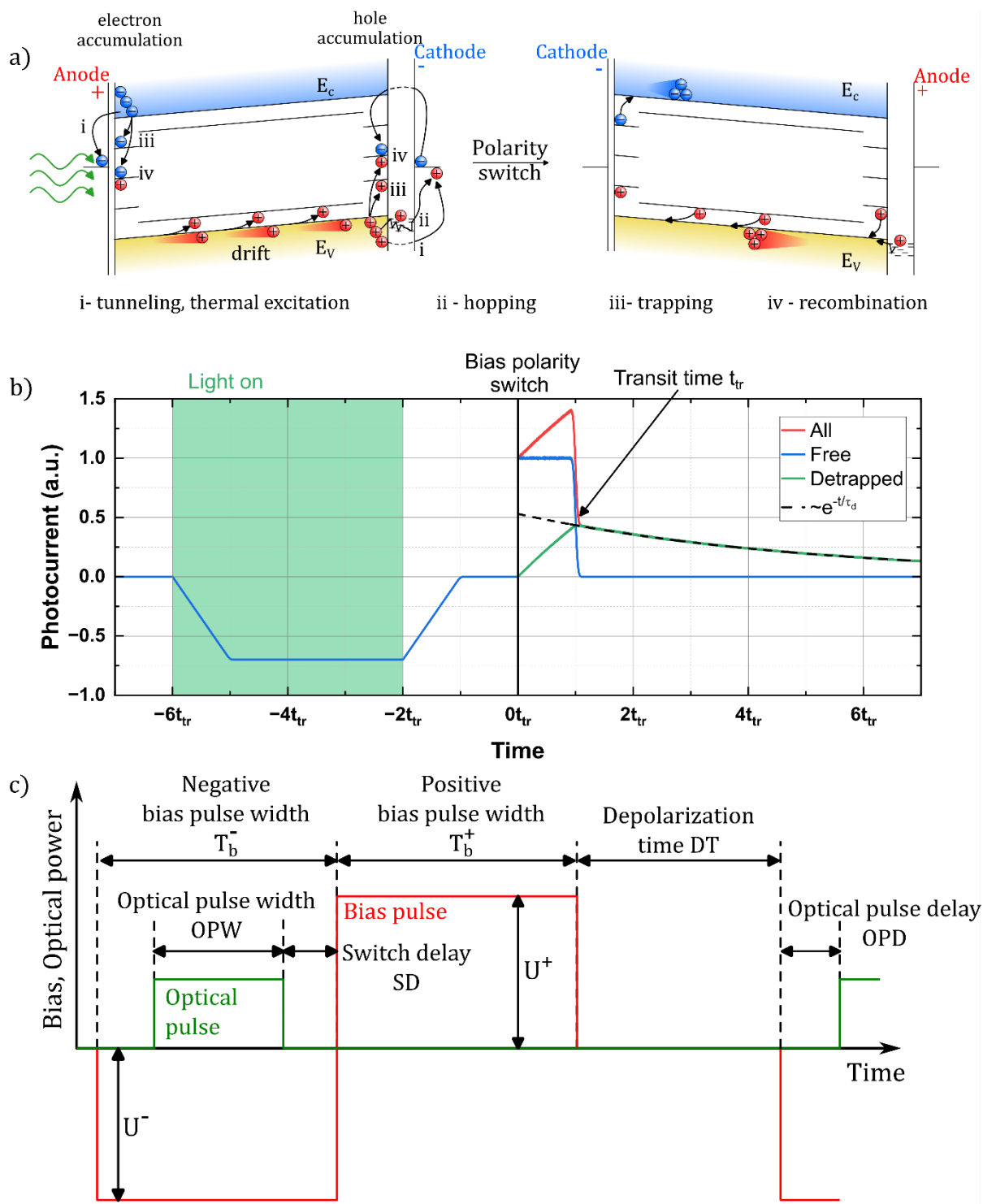

*Fig. 1: a) Principle of BiPS signal generation. Photogenerated free holes dissipate by i) tunneling through or thermal excitation over the barrier, ii) hopping through the barrier or surface layer defects, iii) trapping, or iv) recombination in the surface defects. b) Simulated photocurrent induced by free, detrapped holes and sum of the currents in a sample with uniform electric field. Transit time $t_{tr}$ represents the time it takes charge carriers to drift through the*

*whole sample – transit time. Dashed line represents the emptying of the defect level with detrapping time $\tau_d$ c) Timing diagram of the bias and optical pulses at the BiPS experiment.*

Upon fast bias polarity switch (~100 ns), the accumulated free holes drift towards the cathode (previously anode), inducing a transient signal as shown in Fig. 1b. Because of the strong confinement of accumulated free holes, this component is analogous to a time-of-flight measurement and provides information about the bulk carrier lifetime and the internal electric field profile. Simultaneously, holes that were previously trapped in surface defects are released at various rates, contributing to a long tail of the transient response. Initially, between 0 and $t_{tr}$, the current increases as detrapped holes begin to contribute but have not yet reached the collecting electrode. Only after $t_{tr}$ does the signal reflect the progressive emptying of surface traps as these carriers are collected. Since carrier detrapping is nearly bias-independent, the tail is relatively insensitive to the bias changes and serves as a probe of the surface trap states. Additional experimental and technical details are provided in the Experimental Section at the end of this article. A detailed mathematical description of the BiPS signal generation and modelling, including carrier dynamics and electric field effects, is provided in section SI-3 of the supplementary information (SI).

While BiPS can, in principle, be used without illumination to study dark carrier accumulation, optical excitation offers several benefits: higher signal-to-noise ratio, a well-defined number of generated carriers, and the ability to subtract the capacitive RC response after the bias polarity switch as seen Fig. S4.1. Moreover, thanks to the tunable parameters of the optical pulses (optical pulse delay shown in Fig. 1c it is possible to investigate defect state reoccupation and the effect of the applied bias itself.

## III Electrical field distribution and surface to bulk properties in halide perovskite optoelectronic devices

Building on the photogenerated carrier accumulation, we now turn to a detailed investigation of how electric fields and trap states are distributed across the depth of halide perovskite devices. We studied the bias dependence of BiPS waveforms (CWFs) in CsPbBr$_3$ and MAPbBr$_3$ single crystals to determine the drift mobility of accumulated carriers. Details of the sample preparation, including contact configurations and transport layers, are provided in the Materials section. The initial negative bias pulse ($U^- = -150$ V, $T_b^- = 1$ ms) and the timing of the anode illumination, were kept constant throughout the experiment, while the amplitude $U^+$ and duration $T_b^+$ of the following positive pulse were varied (see timing diagram in Fig. 7a). The obtained BiPS waveforms are presented in Fig. 2a and Fig 2b. The waveforms measured in MAPbBr$_3$ coincide well with the schematic CWF plotted in Fig. 1b. In contrast, the CsPbBr$_3$ waveforms show clear deviations. An initial peak appears immediately after the polarity switch, indicating the presence of negative space charge in a region of the sample adjacent to the anode (see Section SI-3). Conversely, the convex shape of the CWF with a pronounced peak at the transit time $t_{tr}$ suggests the formation of positive space charge near the cathode. As the applied bias increases, the transit time shortens and the current amplitude rises, consistent with an increase in hole drift velocity. The scaling of $t_{tr}$ with bias further supports the time-of-flight-based interpretation. Meanwhile, the long tail observed for $t > t_{tr}$ remains nearly bias-independent, which agrees with hole detrapping from surface states.

The bias normalization in Fig. S4.2 also reveals an unusual increase in the number of the free holes in CsPbBr$_3$, despite keeping the same conditions before the bias polarity switch. Moreover, the total collected charge remains nearly constant, i.e. the higher collected charge induced by free holes is compensated by the reduced collected charge induced by detrapped holes. This observation suggests that the hole detrapping is accelerated by Poole-Frenkel effect[32,33], where the electric field $E$ effectively lowers potential barrier of the defect by $\delta E_t$ given by:

$$\delta E_t = \sqrt{\frac{e^3 E}{\pi \epsilon_0 \varepsilon_{CsPbBr_3}}} \quad ^{32} (1)$$

The barrier lowering increases the number of free or fast-detrapped holes that contribute to the transient signal immediately after the polarity switch. Considering the relative permittivity of CsPbBr$_3$ $\varepsilon_{CsPbBr_3} = 16.4$ [34] the Poole-Frenkel model predicts up to a 20% increase in detrapping rates at electric fields of 1 kV·cm$^{-1}$. A more accurate description would require knowledge of the defect's microscopic nature and falls within the scope Stark effect-induced energy level splitting and shifting rather than purely Coulombic barrier lowering.

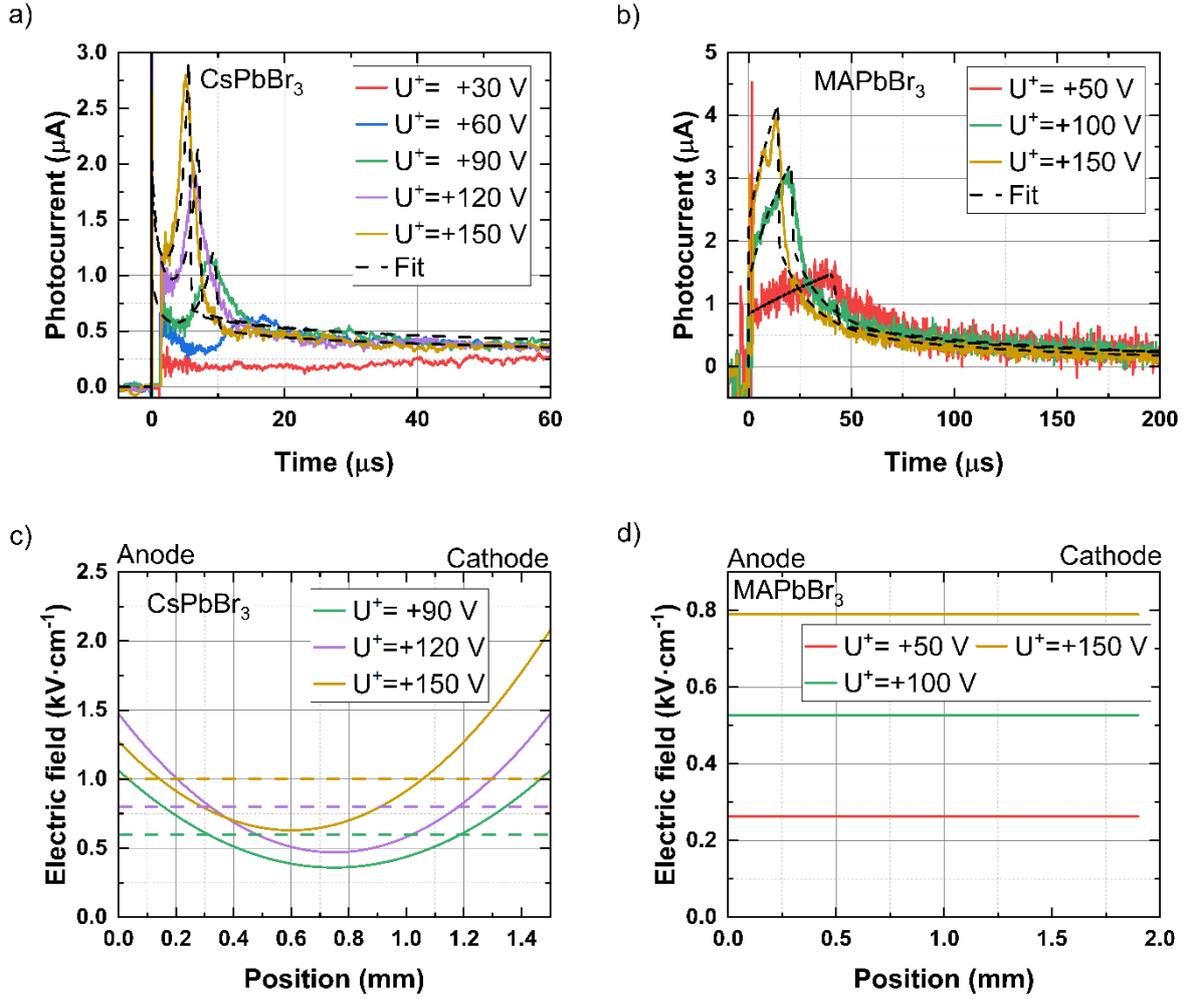

*Fig. 2. Bias dependence of BiPS response and internal field profiles. Measured BiPS waveforms in (a) CsPbBr₃ and (b) MAPbBr₃ single crystals under increasing reverse bias U⁺, overlaid with theoretical fits (dashed lines). Reconstructed internal electric field distributions in (c) CsPbBr₃ and (d) MAPbBr₃ samples. Dashed lines in (c) indicate the steady-state field in the absence of space charge.*

Since the presence of space charge distorts the internal field, the classical formula $\mu = \frac{L^2}{U\, t_{tr}}$ underestimates the true mobility. To determine a precise value, it is necessary to distinguish the contribution of free carriers from the BiPS CWF and to account for the actual profile of the internal electric field. To achieve this, an iterative fitting method described in section SI-3, was used. The best fits are plotted by dashed lines Fig. 2a and Fig. 2b. The obtained mobilities were $\mu_{CsPbBr_3}^{BiPS} = 29 \text{ cm}^2 \cdot \text{V}^{-1} \cdot \text{s}^{-1}$ and $\mu_{MAPbBr_3}^{BiPS} = 14 \text{ cm}^2 \cdot \text{V}^{-1} \cdot \text{s}^{-1}$. These values agree well with those derived from laser-induced transient current technique (L-TCT see SI-2) measured under the same biasing conditions (see SI-2), $\mu_{CsPbBr_3}^{L-TCT} = 29 \text{ cm}^2 \cdot \text{V}^{-1} \cdot \text{s}^{-1}$ and $\mu_{MAPbBr_3}^{L-TCT} =$

15 cm$^2 \cdot$V$^{-1} \cdot$s$^{-1}$. The evaluated hole mobilities are somewhat lower than the commonly reported values of ~50 cm²·V⁻¹·s⁻¹ [35–39] for CsPbBr$_3$ and 24-40 cm²·V⁻¹·s⁻¹ [38,40] for MAPbBr$_3$. This discrepancy may reflect differences in material quality – mainly the presence of shallow traps that lower the observed effective drift mobility. Since we did not see any electron CWFs in L-TCT, even if electrons accumulate in the sample, their contribution to signal is minimal.

Importantly, the absence of a tail in the L-TCT CWFs indicates negligible trapping by shallow bulk defects. This observation is consistent with the concept of defect tolerance in perovskites, which allows for long hole lifetime despite the presence of defects. It thus supports the model in which the tail observed in BiPS CWFs arises primarily from surface defects (see SI-3 for further discussion).

The reconstructed electric fields are plotted in Fig. 2c and Fig. 2d. In MAPbBr$_3$ the electric field is uniform and stable, while in the CsPbBr$_3$ the electric field is significantly warped. As it is evident from Fig. 2c, a region of negative space charge forms beneath the anode (up to 3×10$^{11}$ cm$^{-3}$ see space charge distribution in Fig. S4.2), while a region of positive space charge (up to 3×10$^{11}$ cm$^{-3}$) forms beneath the cathode. Below 60 V, the space charge density is high enough to almost completely screen the electric field in the central region of the sample. In the low-field region (U <60 V), the hole packet significantly broadens due to diffusion and Coulombic repulsion. As a result, the reconstructed electric field represents only an average value over the broadened distribution, rather than a well-defined local field. The presence of low field region also prolongs drift time and smears the falling edge of the BiPS waveform. This effect is highlighted in Fig S4.3 by bias-normalization as a misalignment of CWF falling edges. Under the conditions, where Coulombic repulsion dominates the drift, the nonlinear nature of the system prevents credible distinction of free and detrapped holes. Therefore, CWFs measured at low bias are excluded from electric field evaluation.

While the exact mechanism of space charge formation remains uncertain and requires further investigation, the available evidence strongly supports ionic drift as its primary origin. The long bulk hole lifetime indicates that the space charge is unlikely to arise from trapping on deep levels, but rather from charged shallow defects that do not significantly affect hole dynamics within the timescales of our measurements. Moreover, if we assume that the dark current is dominated by ionic motion, the dark current (Fig. S4.1) integrated over one polarity bias yields the space charge density within the same order of magnitude as the space charge evaluated in BiPS.

## IV. Influence of bias pulse duration on hole accumulation

Because of the use of well-defined photo-excitation, it is useful to normalize the BiPS collected charge by the photogenerated charges. Specifically, we define the normalized fraction of carriers $\eta_t$ as a ratio of the BiPS charge collected up to the integration time t to the total photogenerated charge collected up to the polarity switch, according to eq. 2).

$$\eta_t = \frac{\int_0^t I_{BiPS}(u)du - \int_0^t I_{woBiPS}(u)du}{\int_{-T_b^-}^{0} I_{photo}(u)du} . \quad (2)$$

Here, $I_{BiPS}$ is the photocurrent waveform measured with bias polarity switching, while $I_{woBiPS}$ is the control waveform measured without the polarity switch under otherwise identical conditions. The subtraction in the numerator removes any contribution of carriers that would drift to the electrodes even without polarity reversal, i.e. bulk-trapped carriers slowly released and residual free holes. In practice, such correction requires an additional measurement, therefore the polarity switch delay was chosen sufficiently long ($\geq$ 100 µs, see the photocurrent during illumination Fig. S4.4) so that the second term becomes negligible. The denominator represents the total photogenerated charge collected at the cathode during the initial bias pulse, ensuring that both free and subsequently released holes are consistently taken into account in the normalization. By comparing $\eta_t$ evaluated at the transit time $t = t_{tr}$ ($\eta_{t_{tr}}$) with values for longer integration times $t > t_{tr}$, it is possible to partially distinguish the contributions of free and trapped carriers. While $\eta_{t_{tr}}$ includes both free carriers and those just beginning to be released from traps, the difference $\eta_{t>t_{tr}} - \eta_{t_{tr}}$ consists only from the detrapped holes.

To characterize the effect of the applied bias on the accumulation of free and trapped holes, we studied the hole BiPS signal shown in Fig. 3 as a function of the bias pulse width, while keeping the switch delay constant SD=0.2 ms (see Fig. 7b). This means that only the duration of the applied bias before illumination is varied.

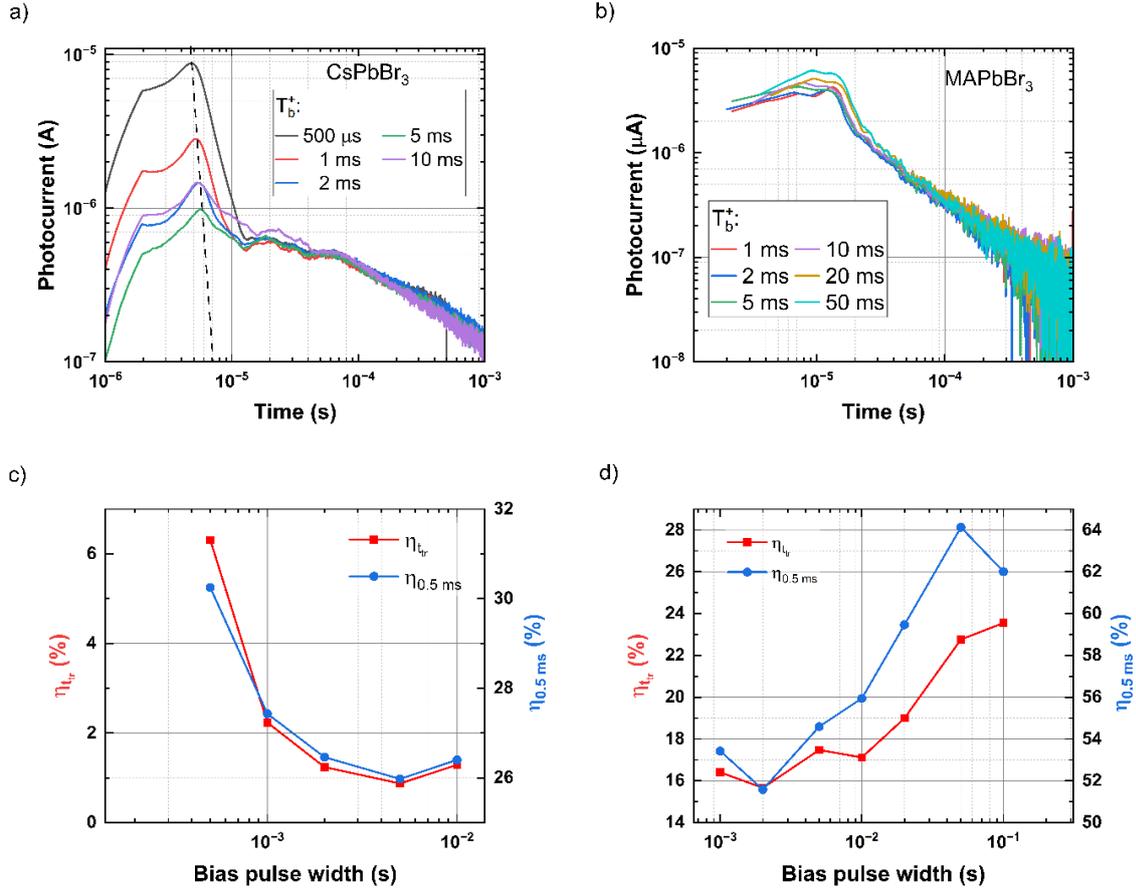

*Fig. 3. Bias-pulse-width dependence of BiPS response and the normalized number of accumulated holes $\eta_{t_{tr}}$ and $\eta_{0.5\,ms}$.* BiPS waveforms in (a) CsPbBr$_3$ and (b) MAPbBr$_3$ single crystals measured at 150 V, for varying bias pulse widths. The dashed vertical line in (a) indicates the hole transit time. Extracted fractions of accumulated free holes and total accumulated charge in (c) CsPbBr$_3$ and (d) MAPbBr$_3$ as a function of bias pulse width.

As shown in Fig. 3c, for the CsPbBr$_3$, $\eta_{0.5\,ms}$ reaches up to 30% of the total photogenerated charge. When $\eta_{t\to\infty}$ is extrapolated using a double-exponential fit of the BiPS tail, the total accumulated charge is estimated to be 60% in CsPbBr$_3$ and 100% in MAPbBr$_3$, confirming that the tail originates from detrapped holes. A comparison of Fig. 3c and Fig 3d reveals a marked difference in the behavior of CsPbBr$_3$ and MAPbBr$_3$ under increasing bias pulse width (BPW). In CsPbBr$_3$, the number of accumulated holes $\eta_t$ decreases with longer BPW, while in MAPbBr$_3$ $\eta_t$ increases. These trends suggest fundamentally different limiting mechanisms in the two materials.

In CsPbBr₃, the shift in transit time (Fig. 3a) indicates that the internal electric field changes during the measurement. This field variation likely accelerates the extraction of free holes by increasing the number of holes that can cross the interfacial barrier (Eq. 1). The ionic drift can also lead to formation of local dipoles between ions and the contact which may become new pathways through barriers or lower the effective barrier height, resulting in more efficient hole extraction.

In contrast, MAPbBr₃ CWFs in Fig. 3b show a stable transit time regardless of BPW. Here, the increase in collected charge is likely due to the progressive filling of deep interfacial states by dark holes. As deep states become saturated more holes remain free $\eta_{t_{tr}}$ or in shallow defects $\eta_{0.5ms}$. The detrapping time of these saturable deep defects needs to be in the range detrapping time 10ms $\ll \tau_d <$ 100ms – long enough not to contribute to BiPS waveform tail but short enough to empty during the depolarization period. Otherwise, the traps with $\tau_d \gg DT$ would become permanently filled and stop affecting the accumulated holes. Moreover, the tail of the BiPS CWFs in Fig. 3b remains nearly constant, suggesting that corresponding levels are mostly full.

Interestingly, Fig. 3c shows that the number of trapped holes in CsPbBr₃ is almost unaffected by the duration of the bias pulse. This observation can be explained by the assumption that only free holes are able to cross the interface barrier while the trapped holes are immobile. Therefore, primarily the free-hole population is affected in CsPbBr₃. In contrast, in MAPbBr₃, the filling of deep surface states (which do not contribute to the tail) shifts the equilibrium and increase both the free-hole concentration and the population of holes trapped in shallow defects.

Finally, the weak increase in collected charge in CsPbBr₃ at very long pulse durations (>5 ms) may be attributed to either return of the electric field toward its original profile (as transit time partially recovers), or to a similar mechanism observed in MAPbBr₃.

## V. Hole extraction dynamics and evaluation of surface defect parameters

Next, we studied the extinction of the accumulated free holes by the prolongation of the bias polarity switch delay SD (see Fig 7c). This is especially useful because hole extinction dynamics contain information about the time evolution of the population of surface defects, and therefore can be used to determine the trapping and detrapping times of surface defects. In this experiment, the bias pulse width was kept the same $T_b^+ = T_b^- =$ 21 ms, only the optical pulse delay OPD and switch delay SD was changed. As can be seen in Fig. 4a, the transit time and therefore the electric field is stable in CsPbBr$_3$. On the other hand, the CWFs in MAPbBr$_3$ in Fig. 4b show a peak near the end of the CWF that vanishes as the photocurrent decreases. Considering the number of accumulated holes, this effect arises from the Coulombic repulsion between free holes rather than from static space charge density associated with trapped carriers or displaced ions[41]. In this case, the rear of the hole packet is slowed down, effectively acting as a space charge that accelerates the front. This results in a rising BiPS waveform with a peak near the end of the CWF and a shift of the peak position, as illustrated in the SI-3. As the number of free holes decreases, the Coulombic repulsion weakens, and the BiPS waveform becomes nearly flat.

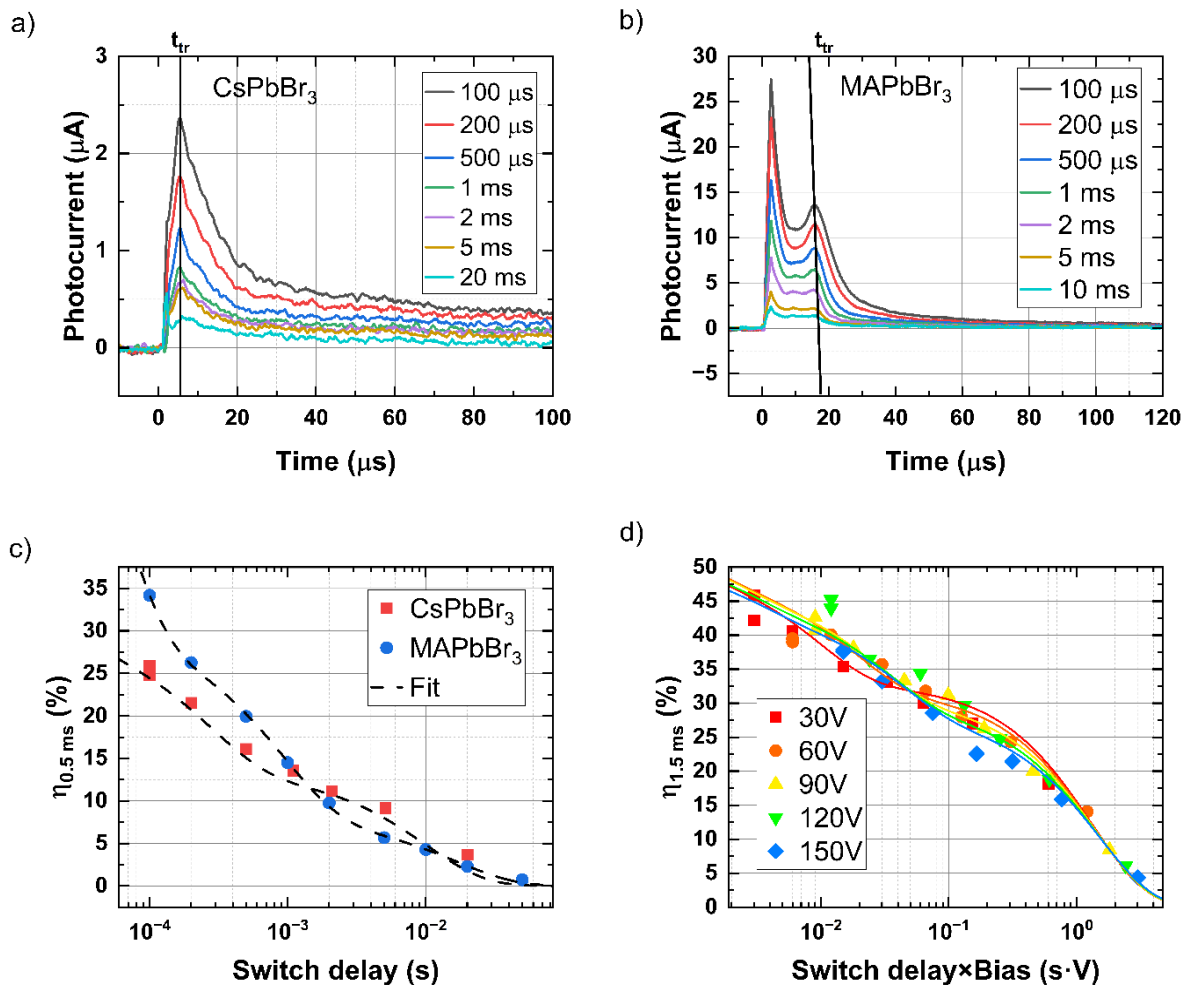

*Fig. 4. Switch delay dependence of BiPS current and accumulated hole density. BiPS waveforms measured at 150 V in (a) CsPbBr$_3$ and (b) MAPbBr$_3$ as a function of the switch delay (SD). (c) Normalized number of accumulated holes $\eta_{0.5\ ms}$ as a function of SD in both samples. (d) Normalized number of accumulated holes at 1.5 ms plotted against SD × bias*

*product for CsPbBr₃. Dashed lines in (c) and solid lines in (d) show fits based on the theoretical model described in section SI-5 of the Supplementary information.*

As can be seen in Fig 4c, the accumulated holes in both samples dissipate with characteristic time in millisecond time scale. Surprisingly, plotting the hole decay in CsPbBr₃ against SD × bias (Fig. 4d) improves continuity and alignment of SD dependencies for all biases. Comparison of the SD and bias×SD dependencies is presented in Fig. S4.5. This suggests that the applied electric field (almost) linearly accelerates the draining of the accumulated holes from the sample. A plausible explanation is that only free holes can overcome the interfacial barrier, and their number at the interface increases linearly with the applied bias, as described by the Boltzmann distribution (Eq.3), assuming non-degenerate statistics and neglecting electrostatic repulsion:

$$p(x) = p_0 \frac{eE_0}{k_B T} e^{-\frac{eE_0}{k_B T}x} \quad (3)$$

Here the interface is located at x=0 (x>0 inside the perovskite), $p_0$ is the total planar density of free holes, $e$ is the elementary charge, $k_B$ is the Boltzmann constant, and $T$ is the temperature. For a 1.5 mm thick perovskite single crystal under 150 V bias, this leads to accumulation of free holes within an approximately 230 nm thick region. Such strong spatial confinement makes BiPS particularly suitable for studying the interface transport phenomena. We assume that the structurally disturbed surface layer (present even after etching) is thicker than this ~230 nm region, so that defects can be considered homogeneously distributed across the probed layer. In this picture, variations in carrier distribution primarily affect transport across the interfacial barrier, while the trapping process itself is governed by the effective density of uniformly distributed surface states. In subsequent analysis, we use this w=230 nm layer thickness to convert volume densities used in the Shockley–Read–Hall model into effective surface defect densities.

To further analyze this effect and estimate defect parameters, we modeled the decay of the BiPS signal using a simplified trapping–detrapping process based on the Shockley-Read-Hall formalism [42,43]. In this approximation, the drift of carriers across the thin probed surface layer is much faster than the characteristic trapping or detrapping times allowing us to omit spatial dependence with the effective surface defect density. This simplification allows fitting with simple set of exponential functions. Although the model neglects trap saturation and detailed spatial distributions, it yields robust order-of-magnitude estimates of the relevant parameters within the accuracy of our experiments. The defects are characterized by their respective trapping $\tau_t$ and detrapping times $\tau_{d0}$, where the detrapping is modified to include the Poole-Frenkel effect $\tau_d = \tau_{d0}\exp\left(-\frac{\delta E}{k_B T}\right)$. Given the variety of mechanisms that may contribute to hole loss at the barrier (see Fig. 1a), the fitting procedure approximates their combined effect using an effective extraction rate at the interface. The bias-scaled extraction rate $\varphi_e$ is defined here as the number of holes removed from the barrier region per unit time and unit voltage; this reflects the total extraction current normalized by the applied bias and is motivated by its approximately linear dependence on voltage observed in Fig. 4d. The whole fitting model is described in SI-5. The parameters obtained from this analysis are summarized in Table 1. From the fits we directly obtain the characteristic trapping times $\tau_t$ detrapping times $\tau_d$ and extraction rates $\varphi_e$ from which other defect capture cross-section and defect ionization energy can be estimated. Because the trapping rate depends only on the product $\sigma_h N_t$, the capture cross-section $\sigma_h$ and defect concentration $N_t$ cannot be separated. To break this ambiguity, we used BiPS waveforms tail to estimate minimum surface defect density to contain all detrapped holes. This value was then used as a lower limit for $N_t$. for each defect. Assuming a hole effective mass of 0.14[44] and a probed surface layer thickness of $w = 230$ nm, we used this $N_t$ to evaluate the corresponding upper limit of the capture cross-section $\sigma_h^*$ and trap ionization energy $E_t^*$.

*Table 1: Evaluated mobilities, hole extraction rates and surface defect level parameters in CsPbBr$_3$ and MAPbBr$_3$ evaluated based on estimated defect density $5\times10^9$ cm$^{-2}$ in CsPbBr$_3$ and $10\times10^9$ cm$^{-2}$ in MAPbBr$_3$ and effective hole mass 0.14.*

| Level | CsPbBr$_3$ | | | | MAPbBr$_3$ | | | |
|---|---|---|---|---|---|---|---|---|
| | $\tau_t$ ($\mu s$) | $\tau_{d0}$ ($ms$) | $E_t^*$ (meV) | $\sigma_h^*$ (cm$^2$) | $\tau_t$ ($\mu s$) | $\tau_{d0}$ ($ms$) | $E_t^*$ (meV) | $\sigma_h^*$ (cm$^{-2}$) |
| 1 | 1.8 | 0.02 | 340 | $4\times10^{-17}$ | 85 | 0.53 | 340 | $2\times10^{-18}$ |
| 2 | 1.3 | 0.27 | 410 | $6\times10^{-17}$ | 140 | 5.7 | 390 | $1\times10^{-18}$ |
| 3 | 0.5 | 3.5 | 500 | $2\times10^{-16}$ | | | | |
| Hole extraction rate $\varphi_e$ | $5\times10^3$ s$^{-1}\cdot$V$^{-1}$ | | | | 21 s$^{-1}\cdot$V$^{-1}$ | | | |
| Hole mobility | 29 cm$^2\cdot$V$^{-1}\cdot$s$^{-1}$ | | | | 14 cm$^2\cdot$V$^{-1}\cdot$s$^{-1}$ | | | |

Based on the parameters summarized in Table 1, the dominant trap level in CsPbBr$_3$ is the deepest one (level 3), which also causes the most significant slowdown in the extraction of holes from the interface. Due to the short detrapping times of the remaining two levels, holes initially trapped in these levels eventually transfer from these shallower states to the deepest one. A similar redistribution of trapped holes appears to occur in MAPbBr$_3$ as well. The estimated capture cross-sections in Table 1 are relatively low, typically on the order of $10^{-17}$–$10^{-16}$ cm$^2$, which aligns with the widely observed defect-tolerant nature of halide perovskites and agrees with values reported in the literature[45,46].

Interestingly, CsPbBr$_3$ shows nearly 250-times faster extraction of free holes compared to MAPbBr$_3$. This difference may be attributed to the presence of the SnO$_2$ transport layer in MAPbBr$_3$, which retains holes more effectively than the Cr contacts used in CsPbBr$_3$.

To further constrain the defect parameters and independently determine the capture cross-section, defect concentration, and activation energy, we performed temperature-dependent BiPS measurements on the CsPbBr$_3$. The goal of this experiment was to extract separate temperature dependencies of the detrapping time and steady-state occupancy, which could resolve the ambiguity between $\sigma_h$ and $N_t$ in the fit. Upon cooling, the extraction of holes through the surface barrier gradually decreased. At 210 K, this led to a pronounced accumulation of photogenerated charge with each optical pulse (Fig. S6.2), resulting in saturation of surface states. The large number of accumulated holes caused the BiPS waveforms to enter a nonlinear regime due to internal screening of electric field, making it impossible to subtract baseline or separate the contributions of free and detrapped carriers. Although this phenomenon prevented more detailed analysis at low temperatures, it is still possible to estimate the maximum concentration of surface defects. From the data in Fig. S6.1, the extracted charge increased by approximately 200 pC after defect saturation. If no hole pass through the barrier, this charge corresponds to the defect-filling capacity of the surface layer. For the estimation, we considered an illuminated spot with a 4 mm diameter and the accumulation layer thickness given by Eq. 3. Although Eq. 3 does not account for Coulombic broadening of the hole packet, it still provides a reasonable upper limit of defect concentration. Under these assumptions, the maximum surface defect density is approximately $1 \times 10^{10}$ cm$^{-2}$, which is consistent with the lower limit of defect concentration assumed in Table 1.

In principle, these limitations could be mitigated by extending the depolarization time and reducing illumination intensity, provided that the signal-to-noise ratio can be improved. Simultaneous measurement of BiPS and TCT waveforms would further simplify the separation of free and detrapped carriers and enable more accurate evaluation of defect population. Such methodological advances would further enhance the robustness of BiPS for characterizing interfacial structures.

VI Conclusion

We developed and applied the Bias-Polarity-Switching (BiPS) technique to investigate charge transport and near-surface defect properties in halide perovskite single crystals. This approach provides access to surface defects located within the first ~200 nm to 2 μm beneath interfacial barriers, while simultaneously probes electric field profile and drift mobility in the bulk of the sample. Importantly, the BiPS methodology and its theoretical description capture the entire process of interfacial carrier dynamics – from the initial accumulation and redistribution during the bias-pulse delay to the eventual release after polarity switching. In contrast to conventional thermo-emission methods, which focus solely on relaxation into equilibrium, BiPS directly links trap filling and detrapping dynamics with the preceding charge-build-up, offering a more complete picture of interfacial processes. It can detect uncompensated bulk space charge as low as ~$10^9$ cm$^{-3}$ in millimeter-thick samples and resolves surface defect dynamics from microseconds to milliseconds. Based on our BiPS measurements, the minimum detectable surface state density is estimated to be on the order of $10^9$ cm$^{-2}$ This combination opens a direct window into dynamic processes that often govern device stability but remain challenging to observe using conventional surface or capacitance-based techniques.

We validated the method on two representative systems: melt-grown CsPbBr$_3$ with symmetric Cr/Cr contacts, and solution-grown MAPbBr$_3$ with an asymmetric Cr, SnO$_2$/Cr. In both cases, BiPS reveals hole-extraction barriers that cause carrier accumulation and populate interfacial trap states. In both samples, we observed interfacial barriers that limited hole extraction and led to the accumulation of photogenerated holes at the contact interface. In CsPbBr$_3$, about 60% of photogenerated holes remain near the contact 200 μs after bias application, whereas in MAPbBr$_3$, almost all photogenerated holes are retained. Moreover, hole extraction in CsPbBr$_3$ is much faster than in MAPbBr$_3$, suggesting that Cr forms lower hole extraction barriers than SnO$_2$.

Using BiPS, we also identified a series of interfacial trap states with trapping time between 1–100 μs and detrapping time ranging from 20 μs to 3 ms, corresponding to activation energies of approximately 300–500 meV with the concentration starting from $10^9$ cm$^{-2}$

In CsPbBr$_3$, we directly identified regions of positive and negative space charge with densities approaching ~$3\times10^{11}$ cm$^{-3}$. This behavior is attributed to ionic drift, which remains inaccessible to standard capacitance-based techniques such as DLCP. Due to similar frequency dependence, DLCP may also misinterpret such space charge buildup as deep trap detrapping.

Our material comparison and methodological framework highlight clear design principles. The lower surface defect density and stable internal field in probed MAPbBr$_3$/SnO$_2$/Cr sample makes it more suitable for optoelectronic applications, while the behavior of CsPbBr3/Cr underlines the importance of improved passivation or alternative contact materials when operational stability is a priority. More broadly, BiPS offers a general platform for probing buried interfaces under realistic working conditions. By quantifying space-charge accumulation, resolving interfacial trap distributions, and disentangling electronic and ionic processes, BiPS links contact design directly to charge extraction and field stability. These capabilities are relevant to the development of more durable perovskite-based solar cells and radiation detectors, and support the transition from lab-scale materials to commercial-grade architectures.

## VII Materials

In this study, we investigated two metal halide perovskite samples: CsPbBr$_3$ crystal was grown by vertical Bridgman technique at Charles University from 5N starting materials CsBr and PbBr$_2$. Cut wafers were mechanically polished by 0.3 μm Al$_2$O$_3$ powder in oil on soft pad, rinsed in toluene, followed by chemical-mechanical polishing in dilute dimethylsulfoxide (DMSO) for 1 minute (30 s each side) on silk and rinsed again in toluene. The final thickness of the wafer used in experiments was L=1.5 mm. Square planar chromium contacts with guard ring structure were prepared by thermal evaporation. The dimensions of the contacts were 7×7 mm$^2$, guard ring width 0.5 mm and guard-ring gap 0.5 mm.

A cuboid MAPbBr$_3$ single crystal was synthesized by inverse temperature crystallization at University of Tennessee[47]. MABr and PbBr$_2$, with a ratio of 1.2: 1, were dissolved in *N,N*-dimethylformamide for a 1 M solution (PbBr$_2$ concentration). The solution was filtered using a 0.2 um PTFE membrane syringe filter. The crystal was grown in a quartz cuvette by gradual heating at a rate of 5 °C/day from room temperature to 70 °C, which was gently rinsed with toluene to remove the residual precursors. The dimensions of the sample were 7.3×6.4 mm$^2$ with thickness L=1.9 mm. The sample was equipped with planar Cr/MAPbBr$_3$/ SnO$_2$/Cr electrodes covering the entire faces of the sample.

Electrical contacts on samples were connected to the sample holder via silver wires and silver conductive paste. To reach good thermal contact with cold fingers, the sample was secured flat to the holder by thermal paste. Mounted samples are shown in Fig.5.

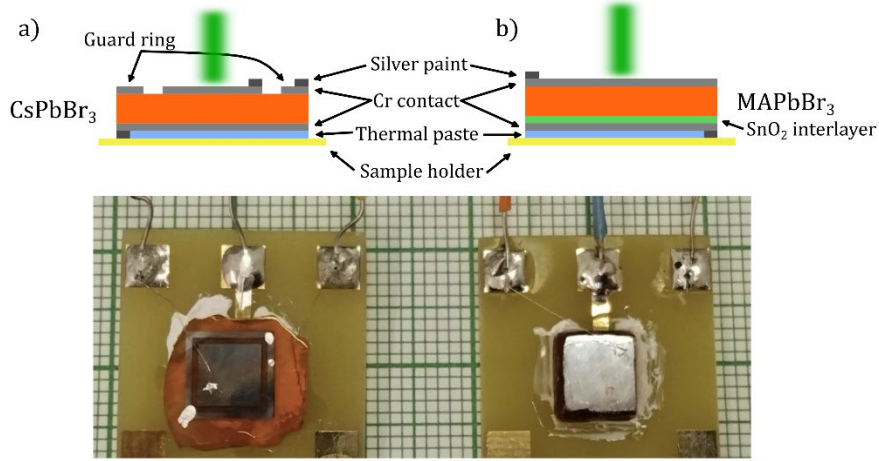

*Fig. 5. Photo of the samples measured in this article. a) CsPbBr3 sample, GR-side up b) MAPbBr3 sample, Cr-side up.*

## VIII Experimental methods

Time evolution of photocurrent after bias polarity switching (BiPS)

For the excitation of CsPbBr$_3$ and MAPbBr$_3$ MHPs, we use a laser diode with wavelength 520 nm (2.38 eV) focused to a 3 mm spot and driven by rectangular current pulses with the width $OPW = 0.1$ ms. Electric current is measured as the voltage drop on 1 kΩ readout resistor and recorded by a digital oscilloscope. The detailed scheme of the setup is shown in Fig. 6.

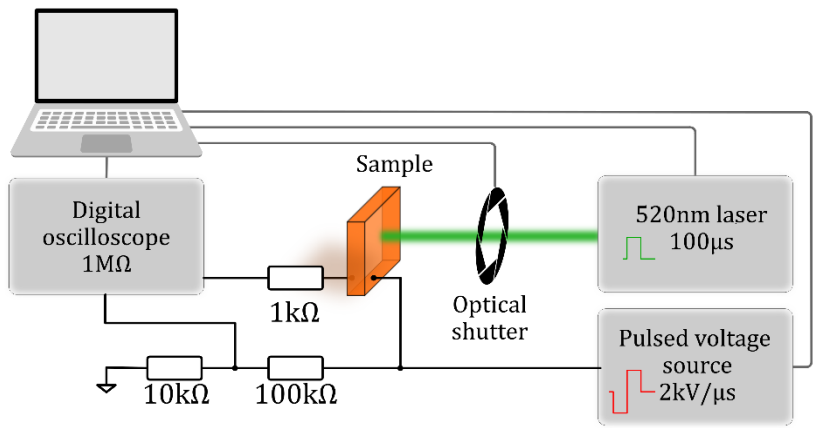

*Fig.6: Schematic diagram of the setup for photocurrent measurement.*

In order to suppress the effects of ionic drift, we use rectangular bipolar bias pulses, displayed in Fig. 1c and Fig. 7, that comply with equal-drift rule $|U^+T_b^+| = |U^-T_b^-|$, where $U^+$, $U^-$ represent the bias amplitudes and $T_b^+$, $T_b^-$ are the bias pulse duration in respective polarities. We consider that this rule ensures that ions return almost to their original positions and slows down the ion-related space charge formation. The bias pulse is followed by a depolarization time DT, when the sample is grounded, that allows defect relaxation (typically ≥200ms). The timing diagram of the bias and optical pulses for individual experiments is shown in Fig. 7. To increase the signal-to-noise ratio, a total of 800–1600 bias cycles were averaged for each bias pulse setting. Data averaging was segmented, where the bias pulse setting was changed every 200 cycles. Finally, each bias setting was repeated 4-8 times over the duration of the measurement in regular intervals. This segmentation allowed monitoring of the potential signal drift during the measurement (see Fig. 8) and ensured that each setting was measured under comparable conditions. The periods indicating significant signal drift, typically the first period, were removed from evaluation, and the rest was averaged. Furthermore, by using an optical shutter, data acquisition alternates between dark and illuminated modes every 20 bias cycles. Photocurrent was evaluated by the subtraction of illuminated and dark measurements in post-processing.

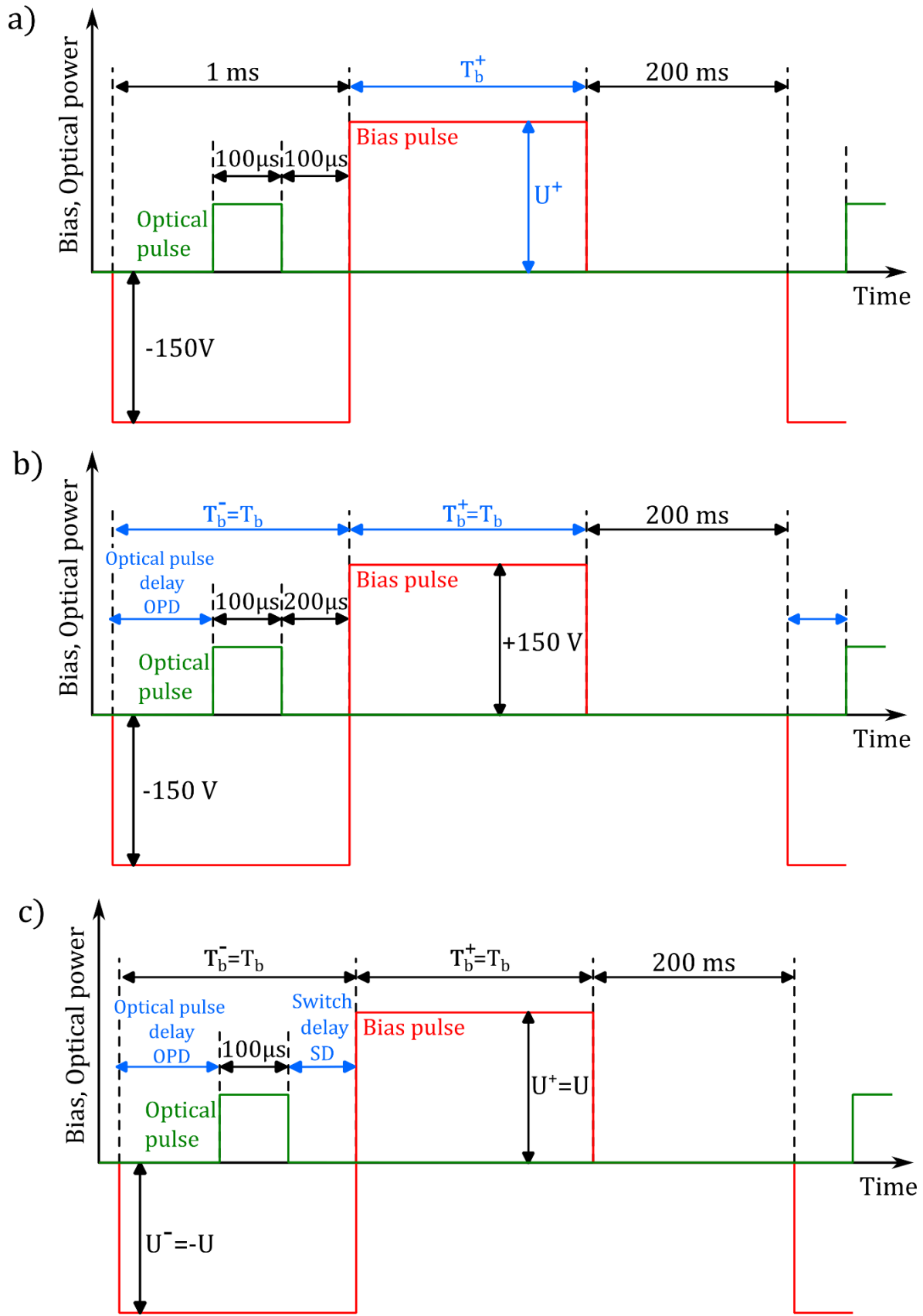

Fig. 7: Experimental parameters used during the measurements of a) BiPS bias dependence in Fig. 2, b) bias pulse width dependence in Fig. 3 and c) switch delay dependence in Fig. 4.

*Parameters varied during experiments are highlighted blue, constant parameters are black. Fig. 7a follows the equal drift rule $|U^+T_b^+| = |U^-T_b^-|$.*

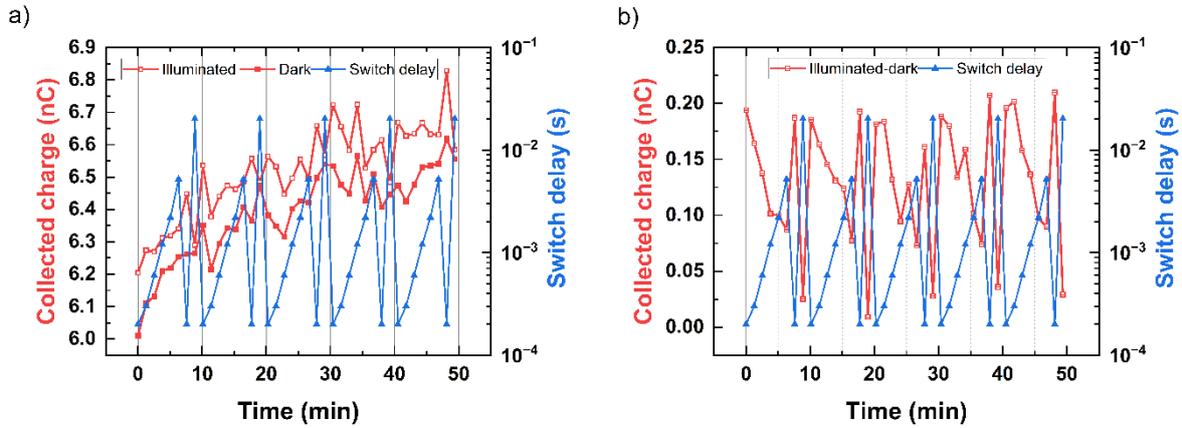

*Fig. 8: Evolution of the collected charge after bias polarity switch over the duration of the experiment.*

## Supporting Information

Additional experimental methods, results, and supplementary figures are provided in the Supporting Information. These include Laser-induced transient current technique (SI-1, SI-2), modeling of the BiPS waveforms and waveform fitting procedure (SI-3), additional BiPS result and temperature dependence measurements (SI-4, SI-6), modeling of the BiPS signal decay (SI-5) and list of abbreviations (SI-7)

## Acknowledgments

This research was supported by the Grant Agency of Charles University under contract No GAUK 393222 and by the Grant Agency of the Czech Republic under contract number 23-07951S. A.M. acknowledges funding from the German Federal Ministry of Education and Research (Bundesministerium für Bildung und Forschung, BMBF) under the NanoMatFutur Call, project number 03XP0625, COMET PV, and the European Union's Framework Program for Research and Innovation HORIZON EUROPE (2021-2027) under the Marie Skłodowska-Curie Action Postdoctoral Fellowships (European Fellowship) 101061809 HyPerGreen. M.A. and J.Y. acknowledge support from the National Science Foundation (NSF), Award Number 2043205, and Micro-Processing Research Facility (MPRF) at the University of Tennessee, Knoxville.

# Microsecond Bias Polarity Switching Reveals Hidden Charge Dynamics at Halide Perovskite Interfaces

Marián Betušiak[1], Roman Grill[1*], Eduard Belas[1], Petr Praus[1], Mykola Brynza[1], Mariselvam Karuppaiya[1], Mahshid Ahmadi[2*], Jonghee Yang[3], Artem Musiienko[4*]

[1] Institute of Physics, Faculty of Mathematics and Physics, Charles University, Ke Karlovu 5, CZ 121 16 Prague, Czech Republic

[*]roman.grill@matfyz.cuni.cz

[2] Joint Institute for Advanced Materials, Department of Materials Science and Engineering, University of Tennessee, Knoxville, TN 37996, USA.

[*]mahmadi3@utk.edu

[3] Department of Chemistry, Yonsei University, Seoul 03722, Republic of Korea

[4] Young Investigator Group, Robotized Optoelectronic Material and Photovoltaic Engineering, Helmholtz-Zentrum Berlin für Materialien und Energie (HZB), Berlin, Germany.

[*]artem.musiienko@helmholtz-berlin.de

### SI-1 Laser-induced transient current technique (L-TCT)

Laser-induced transient current technique with pulsed bias is often used for the characterization of the charge transport and collection efficiency in radiation detectors. It is mainly used for determination of carrier mobility and lifetime [6–9], profile of the internal electric field [7,8,10] and to study the kinetics of the space charge formation [7]. Pulsed bias is used to characterize the transport properties before the space charge forms. In L-TCT, we used the same bias pulses as in BiPS displayed in Fig. 1c). Unlike BiPS, the optical probing is provided by optical pulses with wavelength of 450 nm (2.75 eV) and pulse width of 10 ns. Current induced by photogenerated carriers is amplified and recorded by a digital oscilloscope. Due to spatial localization of a photo-generated charge package, induced current is proportional to the local electric field. It is also is affected by carrier trapping and detrapping. Probing whole bulk of the sample by L-TCT is less convenient for the investigation of surface defect structure which offers only surface recombination velocity derived from the bias dependence of transient current [7,11] Generally, measured transient current is also affected by weighting field[12]. However, in the planar samples investigated in this paper the weighting field is constant 1/L, where L is the electrode separation.

## SI-2 Characterization of bulk transport properties of MHP using the laser-induced transient current technique

Firstly, the L-TCT was used to evaluate basic transport properties of $CsPbBr_3$ and $MAPbBr_3$ samples. In Fig. S2.1a and S2.1b, we present the bias dependence of hole current waveforms (CWFs) measured 0.5 ms after bias application using symetric bias pulses $|U^+| = |U^-|$, $T_b^+ = T_b^- = 1$ ms and depolarization time DT=200 ms. As is evident from the non-constant shape of CWFs in $CsPbBr_3$ sample in Fig. S1 a) a region of negative space charge forms beneath anode (up to -2.5×10$^{11}$ cm$^{-3}$), while a region of positive space charge (up to 10$^{11}$ cm$^{-3}$) forms beneath cathode. This space charge distribution increases the electric field beneath both contacts (Fig. S2.1c) that results in U-shaped CWF. The fact that space charge builds up despite bipolar pulsing suggests that its formation is not completely suppressed this way. However, the use of the bias pulses not complying to the equal drift rule $|U^+ T_b^+| = |U^- T_b^-|$, led to even more significant electric field deformation. Similar and even stronger effects were observed in BiPS experiment. The exact mechanism of space charge formation was not investigated, as it falls outside the scope of this article. Using the Monte Carlo simulation of hole current waveforms, hole mobility $\mu_{CsPbBr_3}^{L-TCT} = 29 \text{ cm}^2 \cdot V^{-1} \cdot s^{-1}$ and electric field profile plotted in Fig. S2.1c) were evaluated.

On the other hand, the hole CWFs in $MAPbBr_3$ in Fig. S1b) are almost constant, and their slope is consistent with uniform electric field Fig. S1d), the hole lifetime of $\tau_{MAPbBr_3}^{holes} = 120$ µs, and the hole drift mobility $\mu_{MAPbBr_3}^{L-TCT} = 15 \text{ cm}^2 \cdot V^{-1} \cdot s^{-1}$.

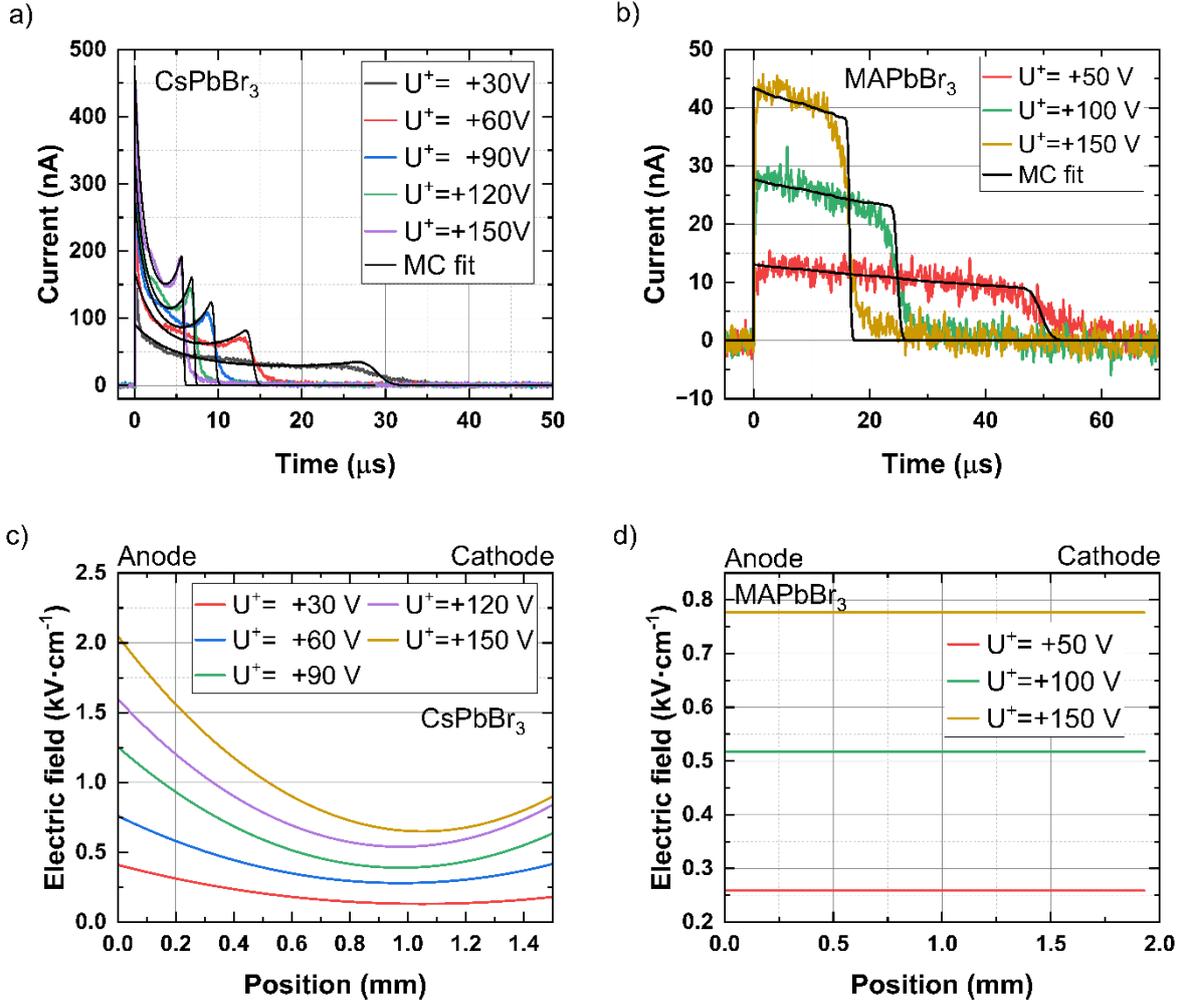

*Fig.S2.1: a) Bias dependence of the hole current waveforms measured in a) CsPbBr$_3$ and b) MAPbBr$_3$ samples by L-TCT fitted by the Monte Carlo simulation (solid black lines). Bias dependence of the internal electric field profile evaluated by Monte Carlo fit of the CWFs in c) CsPbBr$_3$ and d) MAPbBr$_3$ samples.*

### SI-3 Modelling and Analysis of BiPS Transients

According to formula (3) in the main article, holes concentrate in a region thin enough to be treated as a planar source. Assuming that the space charge density (electric field) remains stable in BiPS-CWF time scales then BiPS current transient $I_{BiPS}$ can be expressed in a form of convolution of a single-hole CWF $I_0(t)$ with a charge source representing the hole detrapping from the infinitely thin surface layer (second term in formula SE3.1)

$$I_{BiPS}(t) = S\, p_{free} I_0(t) + S \sum_k \int_0^t \frac{p_k}{\tau_d^k} e^{-\frac{u}{\tau_d^k}} I_0(t-u)\,du \quad \text{SE3.1}$$

where $S$ is the illuminated area, $p_{free}$ is the surface density of the free holes and $p_k$ is the surface density of holes trapped in the *k*-th defect level at the moment of polarity switch and $\tau_d^k$ is the detrapping time of the *k*-th defect level.

At the moment of polarity switch, detrapped carriers do not contribute to the signal, regardless of the electric field distribution or any bulk trapping. The current induced by detrapped carriers increases as progressively more and more detrapped carries drift towards their collecting electrode. At transit time $t_{tr}$, the current reaches maximum and start to fall again as the detrapped carriers are collected. The subsequent decline in current reflects the gradual emptying of surface defect states.

When free carriers are also present and a constant electric field is assumed with negligible bulk trapping, the BiPS current drops at $t_{tr}$ by an amount equal to the initial current. In this case, the contributions of free and detrapped carriers can be clearly separated. However, for non-uniform electric fields, this separation becomes more complicated, as demonstrated by the modeled BiPS-CWFs for constant space charge density shown in Fig. S3.1a and S3.1b. The simulations use a single surface trap with a detrapping time of $\tau_d^{surface} = 100$ μs, a sample thickness of 2 mm, carrier mobility of 25 cm$^2$·V$^{-1}$·s$^{-1}$, and a bias voltage of 150 V. As shown in Fig. S3.1b, the BiPS waveforms are unaffected by the electric field profile after the transit time.

Under these conditions, the contributions of free and trapped holes can be separated using an iterative fitting approach. The method begins with the assumption of a constant electric field, which is used to calculate the expected contribution of detrapped carriers, by fitting the tail of BiPS-CWFs. This component is subtracted from the measured BiPS signal to yield an initial estimate of the free-carrier waveform $I_0(t)$. This estimate is then fitted using Monte Carlo simulations[36,37] to extract the electric field profile, carrier mobility, and diffusion coefficient. The refined field profile is subsequently used for the next iteration. It is important to note that if regions with very low or zero electric field are present, this method becomes unreliable, as the transit time cannot be clearly identified and may be misinterpreted as a contribution of detrapped carriers.

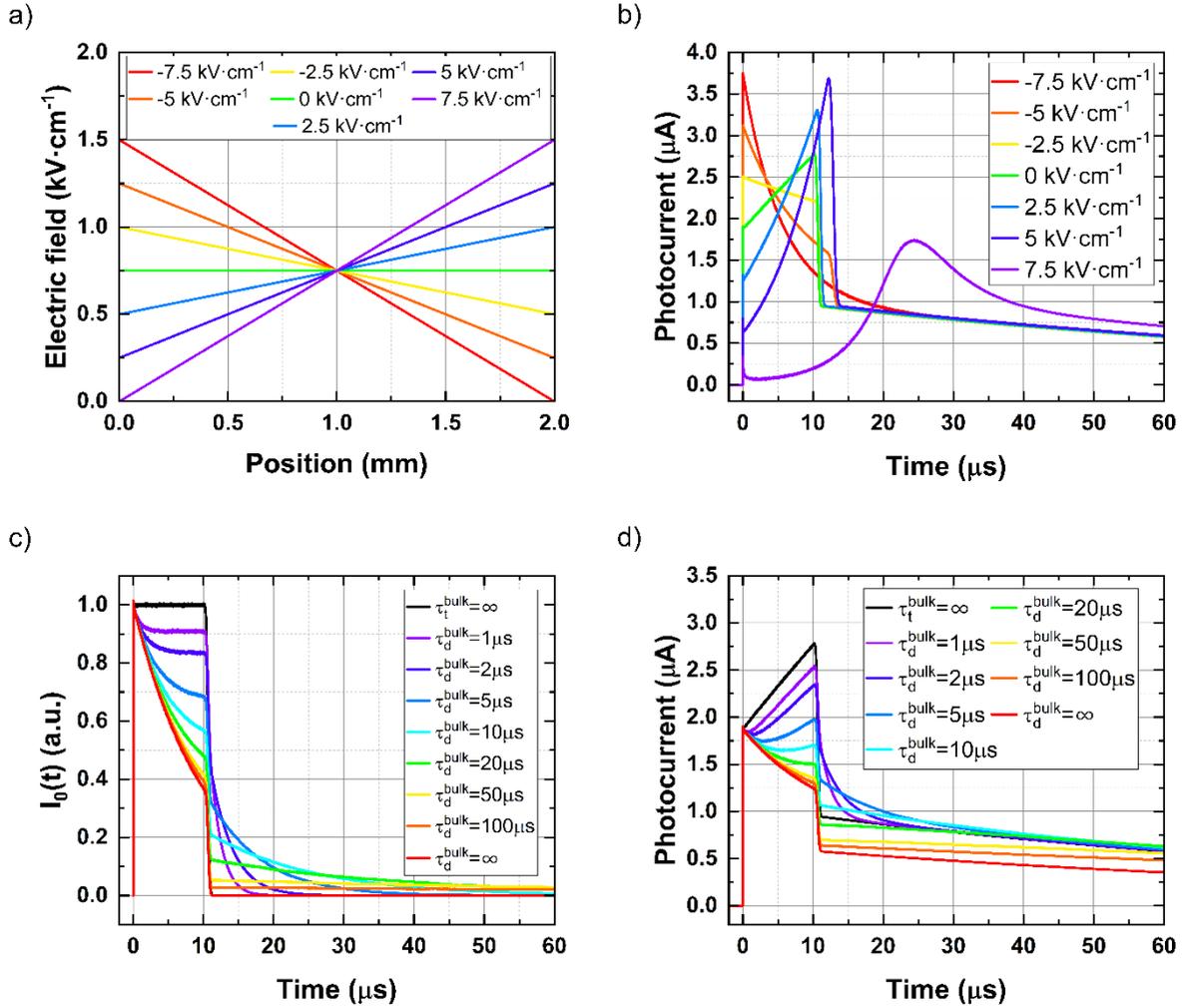

*Fig. S3.1: a) Internal electric field with a constant space charge density and its influence on the shape of the BiPS CWF in b). Fig. c) displays the effect of bulk defects on the initial waveform $I_0(t)$ induced by free carriers. d) The BiPS CWFs corresponding to CWFs in c). In the simulation a single surface defect with $\tau_d^{surface} = 100$ μs detrapping time was used, sample width 2 mm, carrier mobility 25 cm$^2$·V$^{-1}$·s$^{-1}$ and bias 150 V were used.*

On the other hand, bulk defects influence the entire BiPS current waveform. If only deep bulk traps are present and no detrapping occurs (as shown in Fig. S3.1c and S3.1d ), the BiPS tail is uniformly scaled down by a multiplicative factor, while the decay time remains unchanged. This indicates that even in the presence of a non-uniform electric field and finite carrier lifetime, fitting the tail still yields correct detrapping times of surface defects.

However, shallow bulk traps introduce an additional contribution to the tail, which may be misinterpreted as detrapping from surface states—this effect is illustrated in Fig. S3.1d. In such cases, it is necessary to determine the single-carrier waveform $I_0(t)$ from L-TCT measurements. Only with this information can the contributions from surface and bulk defects be reliably distinguished.

Fortunately, as shown in the L-TCT waveforms in section SI-2, CsPbBr$_3$ samples do not exhibit any tail features indicative of bulk trapping, so this complication does not apply here. For the MAPbBr$_3$ sample, however, bulk trapping must be taken into account.

Lastly, for sufficiently high number of accumulated carriers, the drifting carriers distort internal electric field in the sample. This effect can be seen in simulated response in Fig. S3.2 or in measured data in Fig.4b in the main article. Here, the coulombic repulsion slows down the back of the drifting packet and accelerates the front of the packet. This means that part of the carriers reaches the collecting electrode before the expected transit time forming a peak, and the other half is significantly delayed which broadens the falling edge. The curve for 1.2 nC·cm$^{-2}$ correspond to the shortest SD measurement in Fig. 4b).

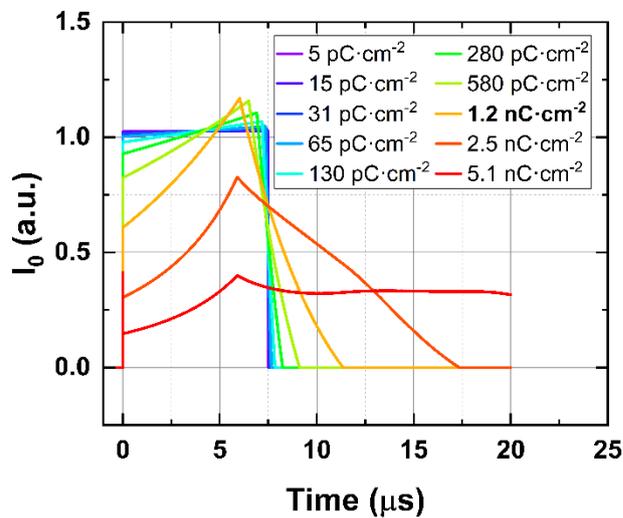

*Fig. S3.2 Effect of Coulombic repulsion on current waveforms for various drifting charge densities. A relative permittivity of 16.4 was used in the 1D simulation of Coulombic repulsion in the electric field 1kV·cm$^{-1}$.*

## SI-4 Additional BiPS results

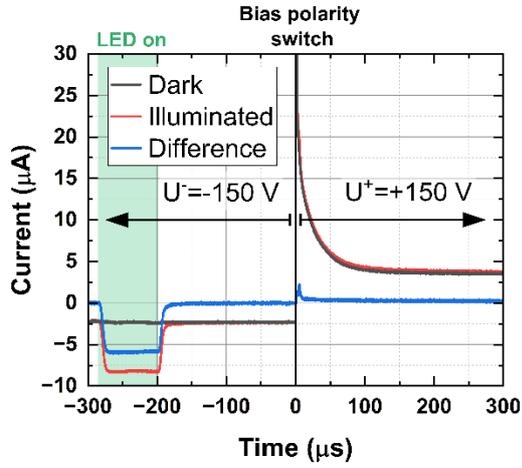

Figure S4.1: Current relaxation in dark and under illumination. Bias pulse parameters $|U^+|=|U^-|=150$ V, $T_b^+=T_b^-=1$ ms, DT=200 ms.

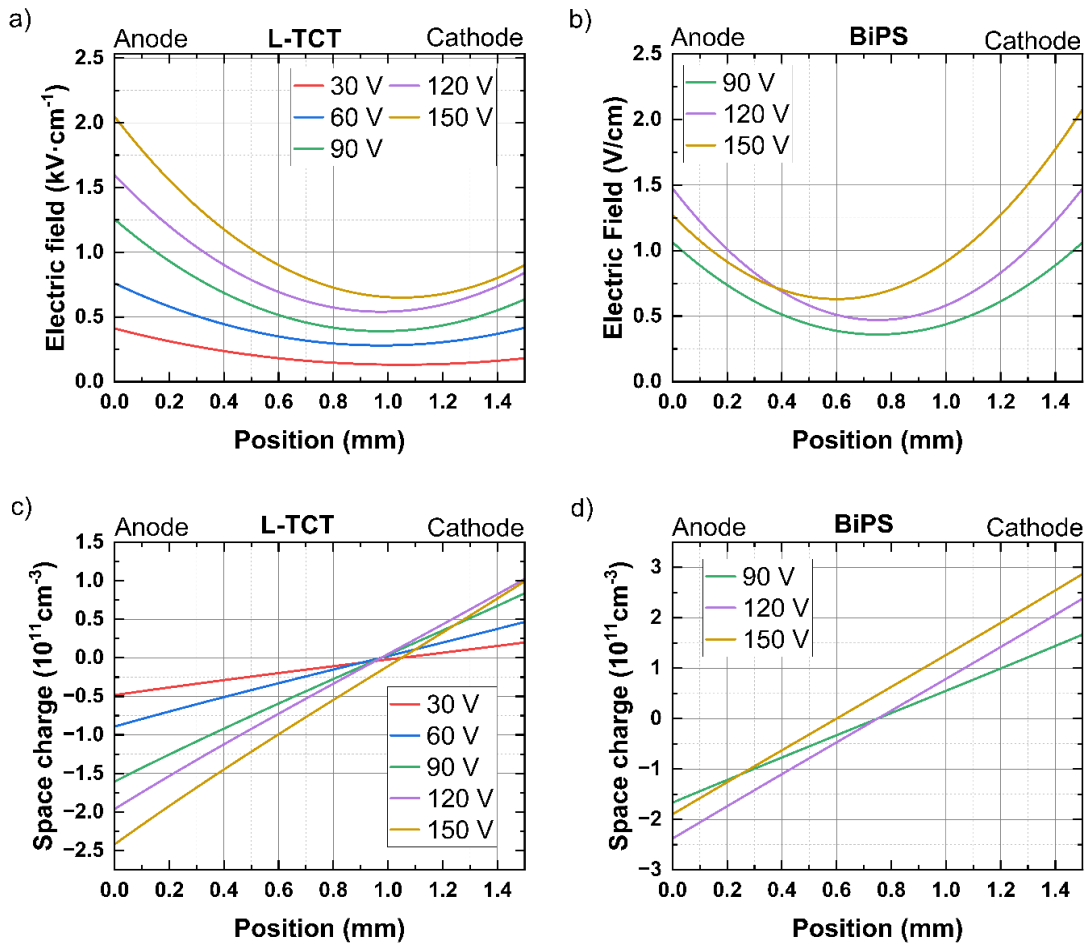

Figure S4.2: Electric field and space charge density evaluated in CsPbBr$_3$ sample using a),c) L-TCT and b),d) BiPS technique.

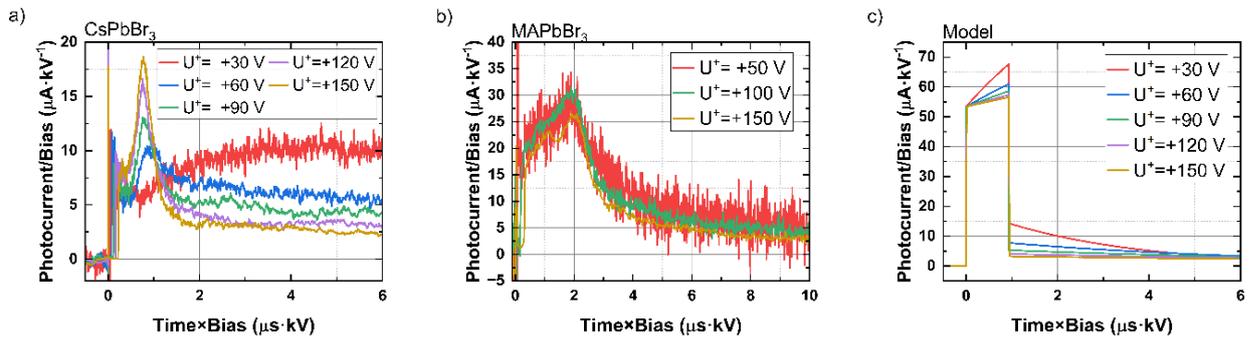

*Figure S4.3: Bias-normalized BiPS current waveforms in a) CsPbBr3, b) MAPbBr3 from Fig.2 in main article and c) modeled signal.*

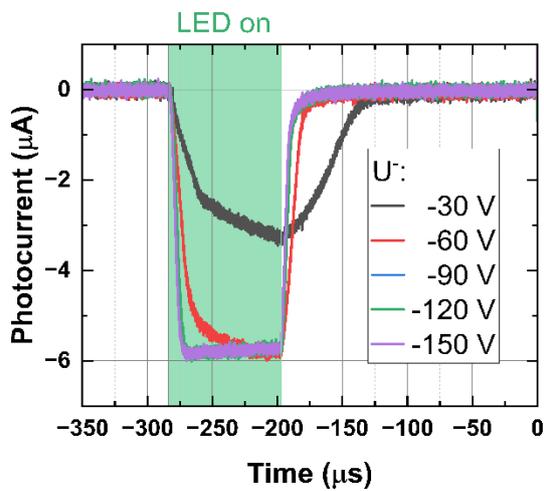

Fig. S4.4: Dependence of pulsed photocurrent used as an excitation in BiPS experiment on the applied bias. Please note that the rising and falling edges change with applied bias.

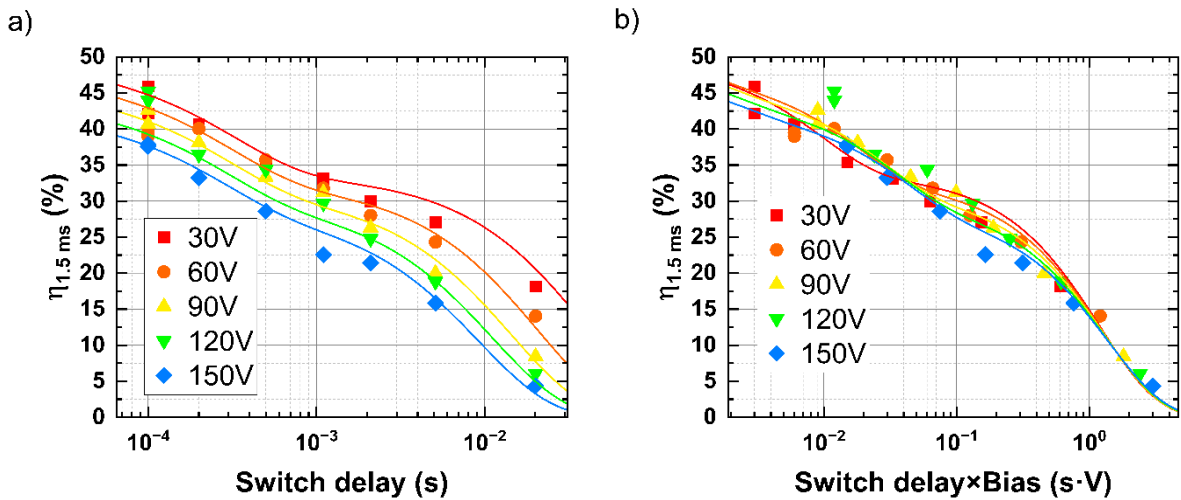

Fig. S4.5: Comparison of the normalized number as a function of a) switch delay, b) switch delay × bias, for various symmetric bias pulses $|U^+|=|U^-|$, $T_b^+=T_b^-=21$ ms, DT=200 ms. Solid line represents the fit by formula (5) in the manuscript.

# SI-5 Modelling of BiPS signal decay and determination of defect parameters

To evaluate defect parameters from the BiPS tail, we adopted a simplified trapping–detrapping scheme based on the Shockley–Read–Hall (SRH) formalism. In this approach, defects are described by their characteristic trapping time $\tau_t$ and detrapping time $\tau_{d0}$, with detrapping modified by the Poole–Frenkel effect:

$$\tau_d = \tau_{d0} \exp\left(-\frac{\sqrt{\frac{e^3 E}{\pi \varepsilon}}}{k_B T}\right)$$

where $\varepsilon$ is the permittivity of the mater, $E$ is electric field in the surface layer, $k_B$ Boltzmann constant a and $T$ is the temperature.

In our experimental conditions, the drift and diffusion time of free holes across the probed surface layer (~230 nm) is much shorter than the characteristic trapping or detrapping times of surface states. As a result, the free-hole redistribute instantaneously allowing us to use the effective density of hole in 0D approximation unless the trap saturation is not explicitly included. Under these conditions the trapping–detrapping dynamics can be reliably captured using the simplified rate-equation model

$$\frac{\partial p}{\partial t} = -\sum_k \tau_t^k p + \sum_k \tau_d^k p^k - \frac{U}{\varphi_e} p \quad \text{SE5.1}$$

$$\frac{\partial p^k}{\partial t} = \sum_k \tau_t^k p - \sum_k \tau_d^k p^k \quad \text{SE5.2}$$

where parameter $\frac{U}{\varphi_e}$ describes bias-dependent extraction rate proportional the number of the free holes directly at the barrier. Assuming roughly rectangular single hole waveform $I_0(t) = \frac{1}{t_{tr}}[\theta(t) - \theta(t - t_{tr})]$ and rectangular excitation optical pulse with the width OPW, then the decay of the BiPS signal $\eta_t(t_{SD})$ at switch delay $t_{SD}$ can be described by

$$\eta_t(t_{SD}) = \mathbf{w}^T \cdot \frac{\mathbf{A}^{-1}}{OPW} \cdot \left(e^{A(t_{SD}+OPW)} - e^{At_{SD}}\right) \cdot \mathbf{p_0} \quad \text{SE5.3}$$

where the matrix **A** defines the coefficients of the system of linear differential equations describing the population of defects in time, $\mathbf{p_0}$ is the initial hole population (all holes are free), and $\mathbf{w}$ is the weight of the contribution to the $\eta_t(t_{SD})$ to reflect the finite integration limits of BiPS CWFs. Defects are described by their respective trapping $\tau_t^k$ and detrapping $\tau_d^k$ times containing Poole-Frenkel shift $\tau_d^k =$

$\tau_{d0}^{k} exp\left(-\frac{\sqrt{\frac{e^3 E}{\pi \varepsilon}}}{k_B T}\right)$ where $\varepsilon$ is the permittivity of the material, while parameter $\frac{U}{\varphi_e}$ describes bias-dependent extraction rate proportional the number of the free holes directly at the barrier.

$$A = \begin{pmatrix} -\frac{U}{\varphi_e} - \Sigma_k \frac{1}{\tau_t^k} & \frac{1}{\tau_d^1} & \frac{1}{\tau_d^2} & \cdots & \frac{1}{\tau_d^k} \\ \frac{1}{\tau_t^1} & -\frac{1}{\tau_d^1} & 0 & \cdots & 0 \\ \frac{1}{\tau_t^2} & 0 & -\frac{1}{\tau_d^2} & \cdots & 0 \\ \vdots & \vdots & \vdots & \ddots & \vdots \\ \frac{1}{\tau_t^k} & 0 & 0 & \cdots & -\frac{1}{\tau_d^k} \end{pmatrix}, \quad p_0 = \begin{pmatrix} 1 \\ 0 \\ 0 \\ \vdots \\ 0 \end{pmatrix}, \quad w = \begin{pmatrix} 1 \\ 1 + \frac{\tau_d^1}{t_{tr}}\left(e^{-\frac{t}{\tau_d^1}} - e^{-\frac{t-t_{tr}}{\tau_d^1}}\right) \\ 1 + \frac{\tau_d^2}{t_{tr}}\left(e^{-\frac{t}{\tau_d^2}} - e^{-\frac{t-t_{tr}}{\tau_d^2}}\right) \\ \vdots \\ 1 + \frac{\tau_d^k}{t_{tr}}\left(e^{-\frac{t}{\tau_d^k}} - e^{-\frac{t-t_{tr}}{\tau_d^k}}\right) \end{pmatrix} \quad \text{SE5.4}$$

Equation SE5.3 can be physically separated into two contributions. The part of the expression containing the matrix **A** describes the time evolution of trap occupancy before the bias polarity switch, i.e. the accumulation and redistribution of carriers at the interface. The weighting vector **w** then projects this evolution onto the detrapping current after the polarity reversal, under the assumption that all released holes drift into the bulk and no retrapping occurs. This separation highlights that the model consistently accounts for both the filling and the emptying stage of the process, in contrast to conventional thermos-emission-type analyses that only evaluate the relaxation step.

## SI-6 Temperature measurements

Figure. S6.1 a) shows the temperature of BiPS CWF for SD=0.1 ms. Temperature dependence could be reliably measured only down to 210 K. Below 190 K, the draining of the accumulated holes became extremely slow and their number increased with each optical pulse despite prolonged depolarization time. This eventually led to accumulation of $2.5 \times 10^{10}$ holes (40 nC) in whole sample over 1 hour measurement at 170 K (Fig. S6.2). In order to explain this phenomenon, we deduce that holes escape the sample by a thermally activated process, which freezes at reduced temperature. This finding rules out barrier tunneling as a mechanism of the holes escape. In addition, also the hole detrapping slows down at the low temperature. Unless, the defects responsible for the BiPS tail are saturable, the $\eta_{1.5\,ms}$ is expected to decrease at low temperature even if the surface barrier is completely impermeable.

As can be seen in Fig. S6.1a, the number of detrapped holes in tail decreases at low temperatures, while the number of free holes increases. This behavior is consistent with the presence of saturable surface defects. From the data in Fig. S6.1, we also observe that the extracted charge increases by approximately 200 pC after saturation, which corresponds to the total filling capacity of the defect-rich interfacial region, assuming that no holes pass through the barrier. This yields a volumetric density of saturable surface defects of ~ $10^{10}$ cm$^{-2}$ (~$4 \times 10^{16}$ cm$^{-3}$), which is consistent with the values reported in Ref. [29]. In Fig. S6.1 b) we present SD dependencies at different temperatures. Temperature dependence of initial values for SD=0.1 ms is caused by the mechanisms mentioned before. At room temperature it behaves as expected. But as the sample cools down, the detrapping damps so the $\eta_{1.5\ ms}$ decreases. This damping is not caused by the escape of holes to the contact. It is due to a slowed detrapping, which extends behind the integration time 1.5 ms. Further reduction of the temperature, below 250K, leads to permanent occupancy of the traps and ensuing significant increase of the free hole lifetime.

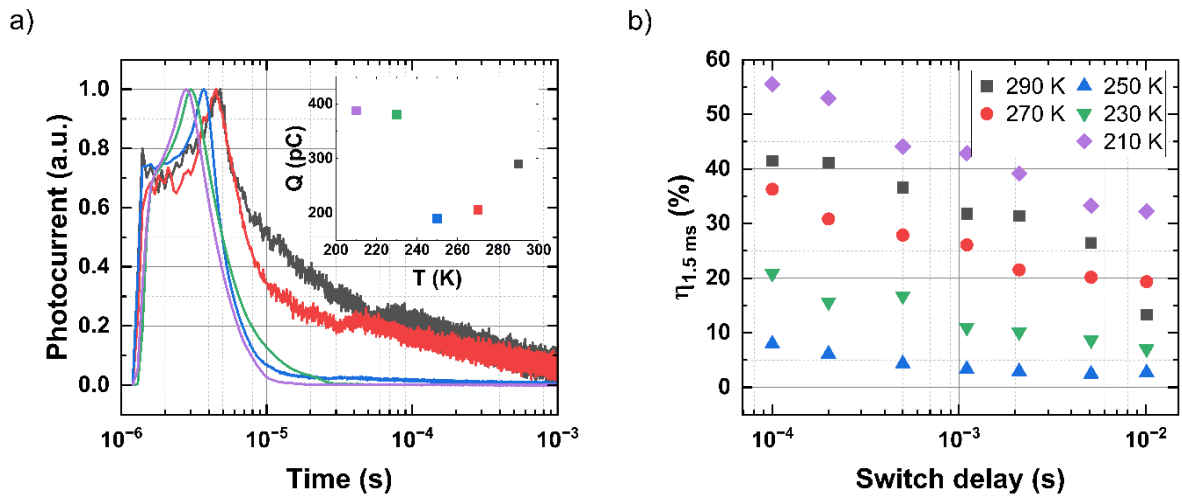

*Fig. S6.1: a) Temperature dependence of normalized BiPS waveforms. Collected charge is shown in the inset. b) Switch delay dependence of the number of the accumulated free holes at different temperatures. Experiment was done at $U^+=U^-=150$ V, $T_b^+=T_b^-=11$ ms, DT=200 ms and switch delay 0.3 ms.*

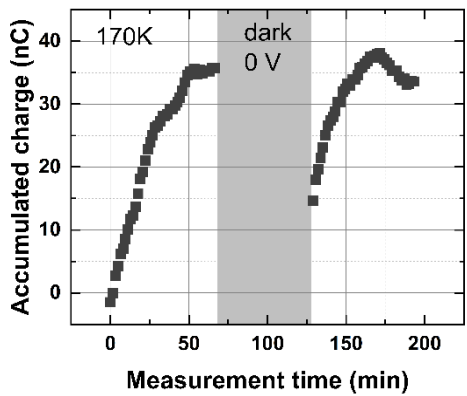

*Figure S6.2: Accumulation of the hole over the duration of the experiments at 170 K. During experiment pulsing parameters $|U^+|=|U^-| = 150$ V, $T_b^+=T_b^-=21$ ms and DT=200 ms were used. During the relaxation period sample was kept in dark under 0 V.*

## SI-7 List of abbreviations

AFM – Atomic Force Microscopy
BiPS – Bias Polarity Switching
BPW – Bias Pulse Width
CWF / CWFs – Current Waveform / Current Waveforms
DLCP – Deep-Level Capacitance Profiling
DT – Depolarization Time
DMSO – Dimethylsulfoxide
EDX – Energy-Dispersive X-ray spectroscopy
KPFM – Kelvin Probe Force Microscopy
L-TCT – Laser-induced Transient Current Technique
MAPbBr$_3$ – Methylammonium Lead Bromide Perovskite
MHPs – Metal Halide Perovskites
OPD – Optical Pulse Delay
OPW – Optical Pulse Delay
PL – Photoluminescence
RC – Resistive-Capacitive (circuit response)
SEM – Scanning Electron Microscopy
SD – Switch Delay
SI – Supplementary Information
$T_b^-$ – width of the negative bias pulse
$T_b^+$ – width of the positive bias pulse
TCT – Transient Current Technique
$t_{tr}$ – transit time
$U^-$ – signed amplitude of the negative bias pulse
$U^+$ – signed amplitude of the positive bias pulse
XPS – X-ray photoelectron spectroscopy